\documentclass{epl}

\title{Statistical Mechanics of an Optical Phase Space Compressor}
\author{Artem M. Dudarev\inst{1,2} \and M. Marder\inst{1,2} \and Qian Niu\inst{1} \and Nathaniel J. Fisch\inst{3}\and Mark G. Raizen\inst{1,2} }
\institute{
  \inst{1} Department of Physics - The University of Texas,
Austin, Texas 78712-1081\\
  \inst{2} Center for Nonlinear
Dynamics - The University of Texas, Austin, Texas 78712-1081\\
  \inst{3} Princeton Plasma Physics Laboratory - Princeton University,
Princeton, NJ 08543
}

\begin{document}

\maketitle

\begin{abstract}
  We describe the statistical mechanics of a new method to produce
  very cold atoms or molecules. The method results from trapping a gas
  in a potential well, and sweeping through the well a semi-permeable
  barrier, one that allows particles to leave but not to return. If
  the sweep is sufficiently slow, all the particles trapped in the
  well compress into an arbitrarily cold gas. We derive analytical
  expressions for the velocity distribution of particles in the
  cold gas, and compare these results with numerical simulations.
\end{abstract}
\section{Introduction} 

Evaporative cooling was originally suggested as a means to achieve
Bose-Einstein condensation  in atomic
hydrogen~\cite{ev-cooling-sug-1,ev-cooling-sug-2,ev-cooling-sug-3}.
Its application to magnetically trapped alkali
atoms\cite{ev-cooling-alk-1,ev-cooling-alk-2} culminated in the first
observation of Bose--Einstein condensation in atomic
vapors~\cite{bec-observation-1,bec-observation-2,bec-observation-3}.
Since then it has been the essential process by which to obtain 
degenerate quantum gases. Nevertheless it has shortcomings. The main two
are
\begin{enumerate} 
\item Almost all atoms originally trapped to produce the condensate
  are lost during the evaporation process.
 \item The time scale for collisions leading to thermal equilibrium
  can be short compared to the time employed to form the condensate. 
\end{enumerate}
 The latter shortcoming is especially severe for fermionic atoms, since 
for two fermions 
in the same state, $s$-wave scattering is forbidden by the Pauli
exclusion principle. Currently, degenerate fermionic gases can only be
obtained by a combination of evaporative and sympathetic
cooling~\cite{symp} in the presence of bosonic atoms or different
states of the fermionic atoms~\cite{ferm-symp}. 

Recently, procedures to construct semi-permeable barriers for ultra--cold
atoms have been suggested~\cite{ours-previous,muga}. Such barriers
transmit atoms coming from one side and reflect them from
another. Their operation relies on different optical shifts for
different internal states. In principle these barriers may be constructed
for many different atoms and molecules. We have shown that by placing such a
wall into a box-shaped potential one may achieve a substantial
increase in phase space density~\cite{ours-previous}, or equivalently,
substantial cooling. 

Our goal in this letter is to demonstrate that by slowly sweeping a
semi--permeable wall through a general trapping potential, the
particles naturally compress into a state of very low energy. The
basic idea is illustrated in Figure \ref{f.1}. At any given time, all
particles remaining in region A to the right of the potential have
energy less than $V(x_b)$. When the wall moves slightly to the right,
the particles that reach it are at their turning point, and have very
small kinetic energy. As the semi--permeable barrier continues to
move to the right, one might think that particles on the left in
region B gain back their energy. However, for a convex potential, as
the particles bounce off the moving wall, they lose more energy in the
collision than they gain otherwise. In this way, a slow sweep of the
semi--permeable barrier through the convex well reduces particle
energies to very low values set by the speed of the sweep.

The conditions needed for this optical compressor to work are quite
different from those required for the effectiveness of evaporative
cooling. The optical compressor tolerates the existence of a
nonequilibrium distribution of particles to the right of the wall. In
particular, the velocities of particles that reach the semi--permeable
barrier from the right are all very low, rather than being given by
the Maxwell--Boltzmann distribution that would describe them in
equilibrium. Thus the process of compression may be fast compared to
the thermal equilibration time of the particles. On the other hand,
the sweep of the wall cannot occur too quickly, because the kinetic
energy of particles after they traverse the wall is given by a
positive power of the wall velocity.

Thus, the optical compressor provides a process completely
complementary to evaporative cooling: 
\begin{enumerate} 
\item No atoms  are lost during the compression process.
 \item The time scale for collisions leading to thermal equilibrium
  must be long compared to the time spent sweeping the
  semi--permeable barrier.
\end{enumerate}

We note that as the equilibration time becomes comparable to the time of the sweep, phase space compression will occur due to combination of evaporative cooling in region A and the process discussed here.

\section{Model}

\begin{figure}
\onefigure[width=7cm]{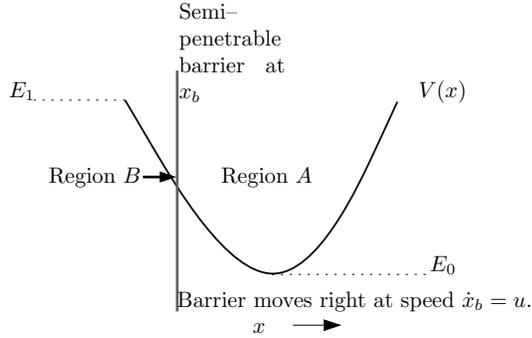}
\caption{Schematic view of the optical compressor. Particles begin in
  region A with characteristic energies $k_BT$. Particles arriving
  at the semi--permeable barrier from the right travel through it,
  while those in region B arriving from the left are reflected. The
  semi--permeable barrier moves to the right at speed $u=\dot x_b$,
  transferring particles from region A to region B, where their
  maximum kinetic energy is much less.
 } 
\label{f.1}
\end{figure}

We consider an ideal collisionless gas trapped in a one-dimensional
potential $V(x)$; two other dimensions are either untrapped or
confined in a box-shaped potential. The gas is originally in thermal
equilibrium at temperature $T=1/k_B \beta$. An ideal infinitely thin
semi--permeable barrier is located at position $x_b$, which is
originally far to the left of any particles.  The barrier moves to the
right with velocity $u=\dot x_b$, its intersection point with the well
moving from $E_1$ to $E_0$ and eventually passing through the whole
well and out the right hand side.

In the limit of slow wall velocities it is possible to obtain
analytical results.  We first focus on the distribution of velocities
with which particles cross the barrier, and then consider the
question of how their velocities change after they have crossed the
barrier.

Particles with energy $E$ are not affected by the wall until the wall
reaches the point where $V(x_b) = E$. Let the period of oscillation of
a particle of energy $E$ in region A be ${\cal T}(E)$. We assume
that there are 
\begin{equation}
n(E)dE=N\beta e^{-\beta E} dE
\label{eq:distribution}
\end{equation}
particles near energy $E$, and that their positions
in the trap are random. Therefore, from the time the first particle of
this energy passes through the barrier, until the last one leaves,
there passes a time ${\cal T}(E)$. The last particle to be captured is
one that had just passed the turning point and was headed to the right
as the barrier reached energy $E$. Particles of energy $E$ will pass
at a uniform rate through the barrier during the time interval ${\cal
  T}(E)$. The first particle to pass the barrier will have no kinetic
energy, while the last one through will have kinetic energy
\begin{equation}
K=-{\partial V\over \partial x_b} u{\cal T}(E)\equiv \dot E {\cal
 T}.\label{eq:last} 
\end{equation}
Here $\dot E$ gives the rate at which the intersection point of the
barrier with the potential well decreases in energy per time. We are
using here the assumption that motion of the semi--permeable barrier
through the well is fast compared to the thermal equilibration time,
or else the kinetic energies of particles escaping the trap would be
described by a Maxwell--Boltzmann distribution with temperature
$T$. We note that even if the semi--permeable barrier moved so slowly
through the trap that thermal equilibrium obtained in region A,
there would still be some cooling in region B, as we now describe. 

Once particles have passed the semi--permeable barrier, they
collide repeatedly with the barrier as it moves to the right and
reflect from it. They lose energy to the barrier in this process. The
final energy of each particle can be determined by observing that the
process is adiabatic in the sense of mechanics, so that the action
$I=\oint pdq$ is conserved\cite{landau-mechanics}.     Consider a
particle that has kinetic 
energy $K$ and total energy $E$ as it passes through the barrier. If
the kinetic 
energy $K$ is not too large, the potential in region B can be
treated as linear, and one computes that the particle has action 
\begin{equation}
  \label{eq:action_after}
  I={2\over 3} {(2mK)^{3/2}\over m \left| V'(E)\right |},
\end{equation}
%
%
where $m$ is the particle mass, and $V'$ is the slope of the 
potential. As the wall continues to move to the right, this action is
preserved, allowing one to determine the final energy $e$
of the particle once the barrier has swept all the way through the
trap. We define in particular the function
\begin{equation}
  \label{eq:func}
 K(e,E).
\end{equation}
which gives the initial kinetic energy $K$ of the particle in terms of its
total final energy $e$, and its initial energy $E$ when it crossed the barrier.

Thus we have the following expression for the distribution of particle
energies $f(e)$ in region B at the end of the compression process:
\begin{equation}
f(e)= \int_{E_{0}}^{E_{1}}dE {dK\over de} \frac{\theta
  (K(e,E) )\theta(\dot{E}\mathcal{T}-K(e,E))N\beta
  e^{-\beta E}}{\dot{E}\mathcal{T}},
\label{eq:dist2}
\end{equation}
Here $\theta$ is a Heaviside step function. 
This expression follows by noting that the $n(E)dE$ particles with potential
energy  $E$ cross the barrier with kinetic energies $K$ evenly distributed between
$0$ and $\dot E{\cal T}$. The energies $E_0$ and $E_1$ are the minimum
and maximum intersection points of the semi--permeable barrier with
the potential well, as indicated in Figure \ref{f.1}. The factor
$dK/de$ accounts for changes in the energy distribution of particles
due to adiabatic expansion in region B.

The distribution $f(e)$ in Eq. (\ref{eq:dist2}) does not describe
thermal equilibrium. Once the compression process has terminated, we
expect that the gas will be maintained for times long compared with
the thermal equilibration time. The total energy of particles in the
trap will be conserved in this process. Thus the end result will be a
thermal distribution of particles with average energy $\bar e_f = E/N$
and temperature $T_f$ that may be found from the system of three
equations with three unknowns (entropy $S$, free energy $F$, and
temperture $T$)~\cite{landau-stat}: 
\begin{eqnarray}
             \nonumber   F &=& - NT\ln \frac{e}{N}\int {\exp\left [{
                -\beta\left( {\frac{{p^2 }}{{2m}} + V(x)}
                \right)}\right ]
                \frac{{dxdp}}{{2\pi \hbar }}} , \\  
                S &=&  - \frac{{\partial F}}{{\partial T}}, \\ 
\nonumber       E &=& F + TS.  
        \label{sys}
\end{eqnarray}
We characterize this final equilibrium distribution by  the efficiency
$\gamma$, defined to be the ratio of phase space density before and
after compression~\cite{psd}:
\begin{equation}
\gamma  = \frac{{\Gamma_f }}{{\Gamma_i }} = \exp \left( {\frac{{S_i  - S_f }}{{k_B N}}} \right).
\end{equation}
Note that for a  power--law potential $V(x) = A x^n$, moving from
initial average energy $\bar e_i$ to final average energy $\bar e_f$
the solution of the system~(\ref{sys}) above gives the compression
\begin{equation}
\gamma  = \frac{{\Gamma _f }}{{\Gamma _i }} = \left( {\frac{{\bar e_i
      }}{{\bar e_f }}} \right)^{\frac{1}{2} + \frac{1}{n}} .
\end{equation}

\section{Examples}

\begin{figure}
        \twofigures[width=7cm]{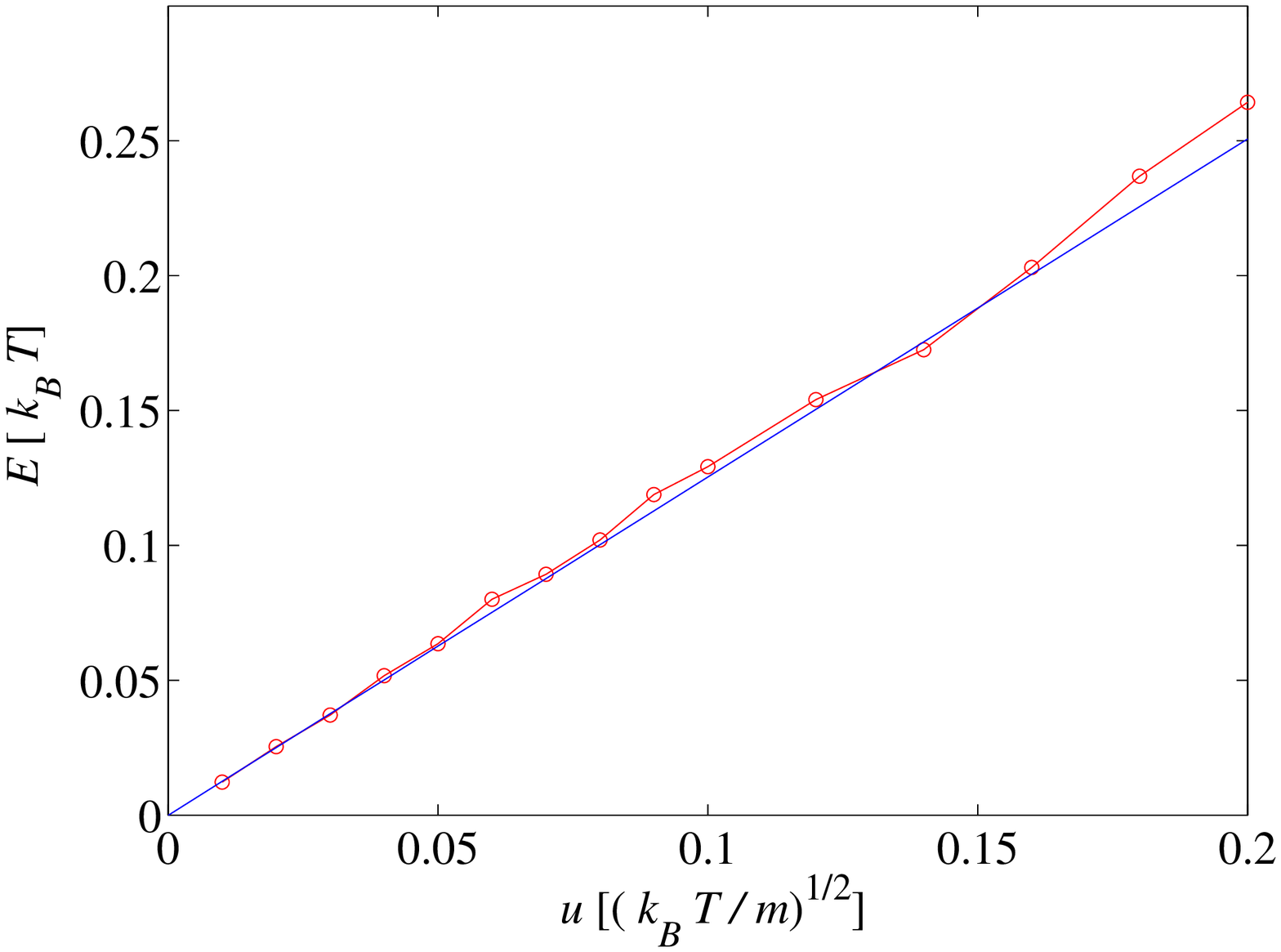}{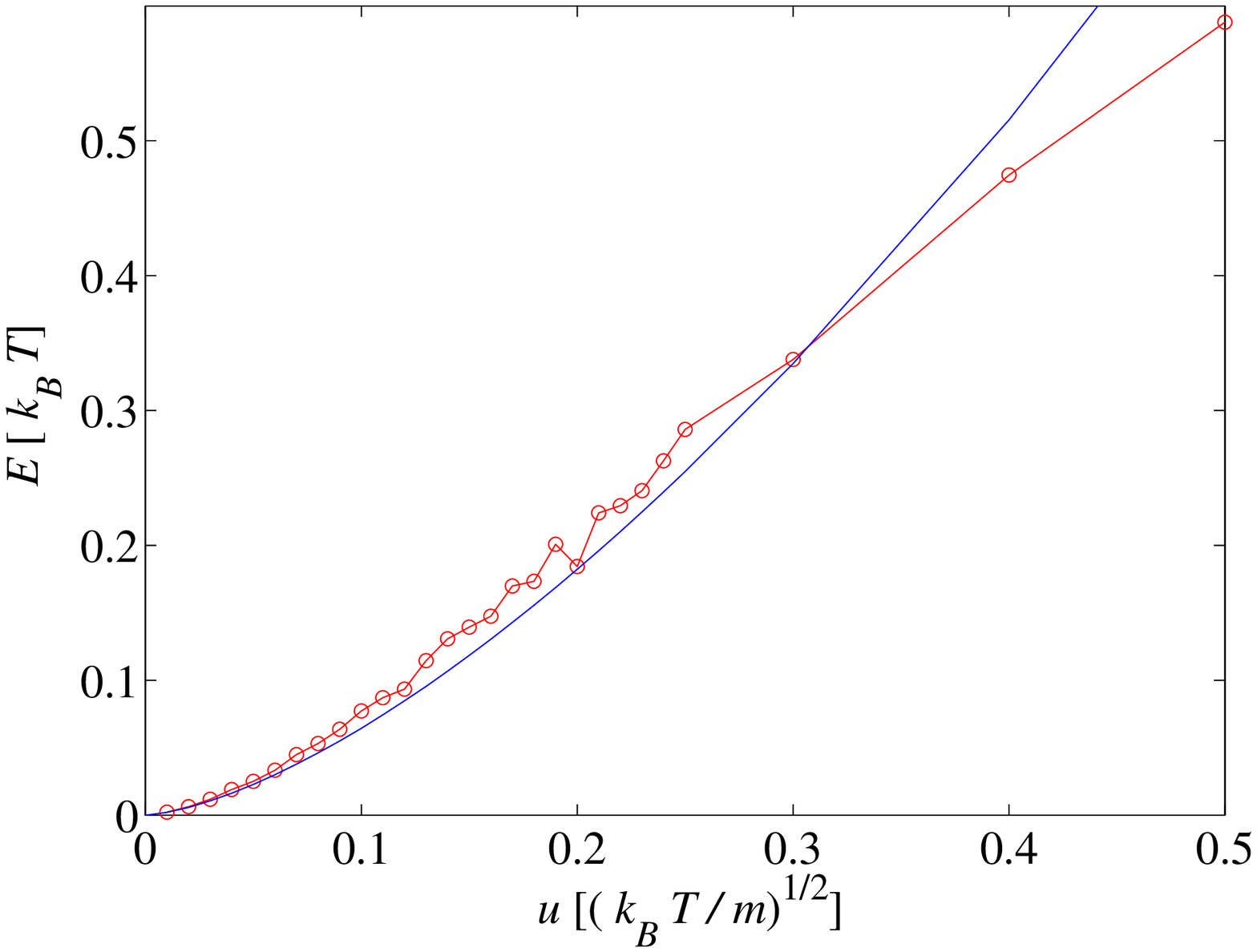}
\caption{Final energy in gravitational trap. The straight line is the
  analytical  result, Eq.~(\ref{eq:ef_grav}). Connected dots come from
  numerical simulations. The wall is initially placed at $E=7 k_B
  T$. Each point  is an average over $N=1000$ particles. \label{f.2} }
\caption{Final energy in parabolic trap. The straight line is the
  analytical  result, Eq.~(\ref{eq:ef_parab}). Connected dots come from
  numerical simulations. The wall is initially placed at $E=3 k_B
  T$. Each point  is an average over $N=1000$ particles.\label{f.3}}
\end{figure}

We now provide examples of two different trapping potentials, and
calculate their effectiveness in cooling dilute gases.

First, consider the \emph{gravitational trap}, defined by
\begin{equation}
  \label{eq:grav}
  V(x)=  \left\{\begin{array}{ll}  -Ax& \mbox{for } x<0\\
        \infty&\mbox{else}.
      \end{array}\right.
  \end{equation}           
As the semi--permeable barrier moves through this potential, the shape of region B does not
change, and therefore the kinetic energy of a
particle when it passes the barrier precisely equals its final total
energy; that is, $K(e,E)=e$.
Carrying out a computation involving the period of
motion in such a potential, we find
\begin{eqnarray}
  f(e) &=& B_1 {\rm erfc} (e/e_0),\\ 
  \mbox{where}\ e_0 &=& 2 \sqrt{2} u \sqrt{m k_B T}.
\end{eqnarray}
Here ${\rm erfc}(x)$ is the complementary error function and $B_1$ is
a normalization coefficient.  From this distribution we obtain the average
energy after compression, 
\begin{equation}
        \bar e_f = \sqrt{\pi/2} u \sqrt{m k_B T}.\label{eq:ef_grav}
\end{equation}
and the efficiency
\begin{equation}
 \gamma = \left( {\frac{{\bar e_i }}{{\bar e_f }}} \right)^{3/2}  = \left( {\frac{9}{{2\pi }}\frac{{k_B T}}{m}} \right)^{3/4} \frac{1}{{u^{3/2} }}.
\end{equation}
The average energy vanishes as velocity of the wall goes to zero.

Next consider the \emph{parabolic trap}
\begin{equation}
  \label{eq:parabola}
  V(x)={1\over 2}A x^2.
\end{equation}
Employing Eq. (\ref{eq:action_after}) we find that
\begin{equation}
        K(e,E) = \left [\frac{3 \pi}{2} \sqrt{E} e\right ]^{2/3}.
\end{equation}
In this case the energy distribution is given by
\begin{eqnarray}
        f(e) = B_2 \left( \frac{e_0}{e} \right)^{1/3} \Gamma \left[ \frac{5}{6}\left( \frac{e}{e_0} \right)^{4}  \right],\\\\ 
        \mbox{where}\ \  e_0 =  \epsilon_0 \left( u \sqrt{\frac{m}{k_B T}} \right)^{3/2} k_B T,
\end{eqnarray}
and $\Gamma[a,x]= \int_{x}^{\infty}dt e^{-t} t^{a-1} $ is an
incomplete Gamma function~\cite{abramowitz}, $B_2$ is another
normalization constant and $\epsilon_0 = 2 \cdot 2^{3/4}(2
\pi)^{3/2}/3 \pi $. The average energy after the process is
\begin{equation}
  \bar e_f = C m^{3/4} u^{3/2} (k_B T)^{1/4}\label{eq:ef_parab}
\end{equation}
where $C = \epsilon_0 \frac{2}{5} \Gamma[\frac{5}{4}] \approx 2.038$.
The efficiency thus depends upon the wall velocity $u$ just as in the
previous example, but with a  different numerical prefactor
\begin{equation}
  \gamma = {\frac{{\bar e_i }}{{\bar e_f }}} = \frac{1}{C}\left(
  {\frac{{k_B T}}{m}} \right)^{3/4} \frac{1}{{u^{3/2} }}. 
\end{equation}

We performed numerical simulations of the process by randomly
preparing particles with various energies in gravitational
and harmonic potentials. We solved the equations of
motion while moving a semi--permeable barrier slowly through the potential.
This procedure was repeated for $N$ particles with average energy
corresponding to the temperature. The results of these simulation are
shown in Figs.~\ref{f.2} and \ref{f.3}. They are in good
agreement with the analytical formulas for small velocities. 

\section{Comparisons and limitations}

Because the one-way wall for an atomic barrier relies upon different
internal states, it truly diminishes the system entropy as a Maxwell
demon would, except for the unavoidable heating due to recoil of a
photon motion.  This can be captured as the cooling effect as
described in this paper.  By comparison, in a plasma, where analogous
one-way walls were proposed in the radio frequency regime~\cite{FDR},
there is no opportunity to change internal states of the plasma ions.
Instead, the one-way ponderomotive-effect wall operates through
Hamiltonian forces only, thereby conserving phase space.  Thus for
plasmas, no matter how the wall is moved, no real cooling can take
place. In the end, if the plasma ions occupy the same volume in space,
they would of necessity occupy the same volume in velocity space --
and hence not achieve a cooling effect.  Note, however, that while the
one-way radio-frequency wall does not cool plasma, it can force ions
or electrons to move in one direction only. Thus, plasma currents can
be driven by plasma waves, which can be useful for a variety of plasma
applications~\cite{RMP}.

The limitation of the semi-permeable wall we suggested 
\cite{ours-previous} is that it results in heating of atoms to a
single photon recoil $m v_r = \hbar k_L$. As the wall velocity diminishes,
the process becomes inefficient. If the temperature of the gas is
originally $n_r$ recoils; i.e. $k_B T = n_r^2 E_r$ where $E_r = \hbar
^2 k^2_L/2/m$, then assuming that the final energy is $E_r$  we find
the slowest velocity with which it is still advantageous to move the
wall in case of the parabolic trap  is
\begin{equation}
        u \approx \frac{0.15}{n_r^{1/3}} v_r.
\end{equation}
In particular, if we start with a temperature of $10$ recoils, the
minimum wall velocity comes out to be $u = 0.05 v_r$. If velocity relaxation
happens on time scale $\tau$, the size of the trap can then be $u \tau$. For
alkalies, $\tau$ can be as long as tens of seconds; hence in this case
the size of the cloud is on the order of centimeters.

\acknowledgments
        MGR acknowledges support from NSF, the R. A. Welch Foundation,
        and the S. W. Richardson Foundation and the US Office of Naval
        Research, Quantum Optics Initiative, Grant
        N0014-04-1-0336. NJF acknowledges support from the US DOE,
        under contract DE-AC02-76-CH03073. MM  thanks the NSF for
        support from DMR-0401766.

\end{document}